\documentclass[12pt]{article}
\usepackage{epsfig}
\newcommand{\nn}{\nonumber}

\newcommand{\be}{\begin{equation}}
\newcommand{\ee}{\end{equation}}
\newcommand{\bea}{\begin{eqnarray}}
\newcommand{\eea}{\end{eqnarray}}

\newcommand{\PCPV}{
\begin{picture}(22,10)
\put(8,-2){\line(2,1){12}}
\put(0,0){$P_{CP}$}
\end{picture}}
\newcommand{\PCPC}{
\begin{picture}(22,10)
\put(0,0){$P_{CP}$}
\end{picture}}

\newcommand{\np}[1]{Nucl.  Phys.  {\bf #1}}
\newcommand{\pl}[1]{Phys.  Lett.  {\bf #1}}
\newcommand{\pr}[1]{Phys.  Rev.  {\bf #1}}

\makeatletter
\begin{document}

\title{\vskip-2.5truecm{\hfill \baselineskip 14pt {{
\small  \\
\hfill MZ-TH/99-56 \\ 
\hfill December 99}}\vskip .9truecm}
 {\bf CP Violation with Three Oscillating Neutrino Flavours
}}

\vspace{5cm}

\author{Gabriela Barenboim\footnote{\tt 
gabriela@thep.physik.uni-mainz.de} 
\phantom{.}and Florian Scheck\footnote{\tt 
Scheck@dipmza.physik.uni-mainz.de}
 \\  \  \\
{\it  Institut f\H ur Physik - Theoretische 
Elementarteilchenphysik }\\
{\it Johannes Gutenberg-Universit\H at, D-55099 Mainz, 
Germany}
\\
}

\date{}
\maketitle
\vfill

\begin{abstract}
\baselineskip 20pt
We explore the prospects of observing leptonic CP violation in a neutrino 
factory in the context of a scenario with three strongly oscillating 
neutrinos
able to account for the solar, the atmospheric and the LSND results.
We address also the problems related with the fake asymmetries induced
by the experimental device and by the presence of matter.
\end{abstract}
\vfill
\thispagestyle{empty}

\newpage
\pagestyle{plain}
\setcounter{page}{1}

\section{Measuring leptonic CP violation with neutrino beams from a
  muon collider}

Muon storage rings, muon colliders, and their physics potential are
being studied intensively at FNAL and at CERN \cite{FLC}. In
particular, a muon storage ring at some 20~GeV, being the first
step in these projects, would serve to produce intense neutrino beams
of unique quality. This possibility  has received much
attention \cite{geer,belen,belen2,rom} recently. The straight sections 
of a muon collider would serve as sources of $\overline{\nu}_\mu$ and 
of $\nu_e$ with  energy spectra perfectly calculable from muon decay,
when positive muons are stored, and, similarly, of $\nu_\mu$ and
$\overline{\nu}_e$  with well-known energy spectra when
negative muons are stored. 

In this work we explore the possibilities of studying
CP~violation in the leptonic sector, at a neutrino factory of this
kind, by comparing the oscillation probabilities of CP-conjugate
channels $\nu_i \rightarrow \nu_j $ and
$\bar{\nu_i} \rightarrow \bar{\nu}_j$ with $(i \neq j)$. The most
suitable channels for studying CP violation are
$\nu_e \rightarrow \nu_\mu $ and
$\bar{\nu}_e \rightarrow \bar{\nu}_\mu$, as well as their T~conjugate partners 
$\nu_\mu \rightarrow \nu_e $ and
$\bar{\nu}_\mu \rightarrow \bar{\nu}_e$. In these channels, unlike the 
$\nu_\mu \rightarrow \nu_\tau $ channels, the CP violating part of the
oscillation probability is not hidden by the CP conserving part
\cite{lindner}, so that large asymmetries between CP-conjugate
channels may arise, provided the leptonic CKM matrix allows for large
violation of CP. Among these, the channels $\nu_e \rightarrow \nu_\mu $ and
$\bar{\nu}_e \rightarrow \bar{\nu}_\mu$ seem to be the most promising
because it may be easier to disentangle negative from positive muons,
in a large detector of high density, than to disentangle electrons
from positrons. 

As a matter of example, we study the case of neutrino beams from
stored muons with energy $E_\mu =20$~GeV, and an experimental
arrangement where they travel over a distance of some 730~km,
i.e. from CERN to the Gran Sasso laboratory, or, likewise, from FNAL
to the Soudan mine. But we also give the scaling 
behaviour of the effects with either energy or distance. Within the
range of squared mass differences and mixing angles, matter effects 
are important, see also \cite{belen}. However, unlike the case where
neutrinos traverse the Earth from the antipode \cite{nosupdown}, they
are easy to cope with because the density in the Earth's crust is
essentially constant. 

The $\nu_e \rightarrow \nu_\mu $ oscillation probability can be
measured as follows: Suppose the electron neutrinos are produced by
the decay of a number $N_{\mu^+}$ of positive muons in the straight
section of the storage ring pointing to  the detector. The $\nu_\mu$
which appear when there is oscillation of $\nu_e$ into $\nu_\mu$,
are detected by their charged current interaction in the detector.
The number of observed muon neutrinos, $n_{\nu_\mu}$ is given by,
\bea
n_{\nu_\mu} = N_{kT}\; 10^9 \;N_A \int \;F_{\nu_e}\; \sigma_{\nu_\mu}
\;  P(\nu_e \rightarrow \nu_\mu) \; dE
\eea
where $F_{\nu_e}$ is the forward flux of electron neutrinos from 
a number $N_{\mu^+}$ of
positive muon decays, $\sigma_{\nu_\mu}$ is the charged current cross
section per nucleon and $ P(\nu_e \rightarrow \nu_\mu)$ is the 
oscillation probability for neutrinos traveling inside the Earth taking
into account matter effects.
$N_{kT}$ is the size of the detector in kilotons. An analogous formula
holds in the case of anti-neutrinos. 

Adopting the sample design configuration for muon production,
cooling, acceleration and storage described by Geer \cite{geer}, the
number of available muons of either sign is approximately 
$ 8 \cdot 10^{20}$ per year, for muons stored at an energy of 20~GeV.
Of these, one fourth decay in a straight section directed towards the
neutrino detector with a 10 kT target some $\sim 730$~km downstream, 
yielding about $ 2 \cdot 10^{20}$ neutrinos per year and an identical number
of anti-neutrinos. We use these numbers in what follows, and refer the
reader to \cite{geer} for details of the design of neutrino beams.

Let us begin by computing the number of produced muon
neutrinos. Experimental cuts needed to eliminate background as well
as detecting efficiencies will be included later on. The neutrino
fluxes at a neutrino factory have simple analytical forms that follow
from the well-known formulae for muon decay. Let $x= E_\nu / E_\mu$ be
the fractional neutrino energy. For unpolarized muons of either
charge, and neglecting corrections of order $m_\mu^2/E_\mu^2$ the
normalized fluxes of forward moving electron neutrinos are
\bea
g_{\nu_e,\bar{\nu}_e}(x) = 12 \; x^2 \; (1-x)
\eea
and, for each neutrino type, the flux in the forward direction due to
$N_\mu$ decaying muons is
\bea
F\;=\;  \left.\frac{d^2  N_\nu}{dx \; d\Omega}\right|_{\theta \simeq 0} \; = \;
\frac{E_\mu^2\; N_\mu}{\pi \;m_\mu^2 \; L^2}\; g_\nu(x)
\eea

The above expressions are valid for a detector placed in the forward
direction whose transverse dimensions are much smaller than the beam's
transverse size $\sim (L\; m_\mu /2\; E_\mu)$.

We assume that the interaction cross sections due to charged current interactions scale linearly with the energy even in the very low energy
part of the spectrum
\bea
\sigma_{\nu_e}= .67 \cdot 10^{-38} \mbox{cm$^2$ $E$ (GeV)} \\
\sigma_{\bar{\nu}_e}= .34 \cdot 10^{-38} \mbox{cm$^2$ $E$ (GeV)} 
\eea

Regarding matter effects, let us remind the reader of the fact that of
all neutrino species only $\nu_e$ and $\bar{\nu}_e$ have elastic
scattering amplitudes on electrons due to charged current interaction.
This, as is well known, induces effective ``masses'' $\mu = \pm 2
E_\nu a $, where the upper sign refers to the electron neutrino, the
lower sign to the corresponding anti-neutrino, and where $a = \sqrt{2}
G_F n_e$, $n_e$ being the electron density.

Matter effects \cite{mat} are important provided the interaction term
$\mu$, 
\bea\label{matter}
\mu = 7.7 \cdot 10^{-5} \mbox{eV}^2 
\left(\frac{\rho}{ \mbox{gr/cm$^3$}}\right) 
\left(\frac{E_\nu}{\mbox{GeV}} \right)
\eea
is comparable to, or bigger than, the
quantity $\Delta_{m_{ij}^2}= m_i^2-m_j^2 $  
for some mass difference and neutrino  energy.

CP related observables often involve the comparison between 
measurements in two charge-conjugate modes of the factory. 
One example of an asymmetry is \cite{cab}
\bea
a^{tot}_{CP} = \frac{ \int P(\nu_e \rightarrow \nu_\mu) F_{\nu_e} \sigma_{\nu_e} dE -
 \int P(\bar{\nu}_e \rightarrow \bar{\nu}_\mu) F_{\bar{\nu}_e} 
\sigma_{\bar{\nu}_e} dE }
{ \int P(\nu_e \rightarrow \nu_\mu) F_{\nu_e} \sigma_{\nu_e} dE +
 \int P(\bar{\nu}_e \rightarrow \bar{\nu}_\mu) F_{\bar{\nu}_e} 
\sigma_{\bar{\nu}_e} dE }
\eea
or in other terms,
\bea
a^{tot}_{CP}= \frac{n_{\nu_\mu}/N_{\mu^+}\; - \; n_{\bar{\nu}_\mu}/N_{\mu^-}}
{n_{\nu_\mu}/N_{\mu^+} \;+ \; n_{\bar{\nu}_\mu}/N_{\mu^-}}
\label{stat}
\eea
In vacuum this quantity $a^{tot}_{CP}$ would be a pure CP odd
observable. The voyage through our CP uneven planet, however, induces
a nonzero asymmetry even if CP is conserved, since $\nu_e$ and
$\bar{\nu}_e$  are affected differently by the electrons in the Earth
\cite{ar}. Therefore, to obtain the genuine CP odd quantity of
interest, the matter effects must be subtracted with sufficient 
precision.

For this purpose, we compute the matter asymmetry in the absence
of CP violation, or fake CP asymmetry, by 
\bea\label{aCPzero}
a_{CP}(\delta =0 )  
= \frac{ \int P(\nu_e \rightarrow \nu_\mu)\mid_{\delta=0} \;
 F_{\nu_e} \sigma_{\nu_e} dE 
\;  - \;
 \int P(\bar{\nu}_e \rightarrow \bar{\nu}_\mu)\mid_{\delta=0} \; 
F_{\bar{\nu}_e} 
\sigma_{\bar{\nu}_e} dE }
{ \int P(\nu_e \rightarrow \nu_\mu)\mid_{\delta=0} \;
 F_{\nu_e} \sigma_{\nu_e} dE 
\;  + \;
 \int P(\bar{\nu}_e \rightarrow \bar{\nu}_\mu)\mid_{\delta=0} \; 
F_{\bar{\nu}_e} 
\sigma_{\bar{\nu}_e} dE }
\eea
where we take into account matter effects but set $\delta = 0$ 
in the transition probabilities.

The total asymmetry $a^{tot}_{CP}$ that will be found in an experiment of
the type described above, is a function of $a_{CP}(\delta = 0)$,
eq.~(\ref{aCPzero}), and of the asymmetry in vacuum (taking due account
of CP violation)
\bea
a_{CP}^{vac} = \frac{ \int P^{vac}(\nu_e \rightarrow \nu_\mu) \;
 F_{\nu_e}\; \sigma_{\nu_e}\; dE 
\;  - \;
 \int P^{vac}(\bar{\nu}_e \rightarrow \bar{\nu}_\mu) \; 
F_{\bar{\nu}_e}\; 
\sigma_{\bar{\nu}_e}\; dE }
{ \int P^{vac}(\nu_e \rightarrow \nu_\mu) \;
 F_{\nu_e}\; \sigma_{\nu_e}\; dE 
\;  + \;
 \int P^{vac}(\bar{\nu}_e \rightarrow \bar{\nu}_\mu) \; 
F_{\bar{\nu}_e}\; 
\sigma_{\bar{\nu}_e}\; dE }
\eea
where $ P^{vac}(\nu_e \rightarrow \nu_\mu)$ and $P^{vac}(\bar{\nu}_e
\rightarrow  
\bar{\nu}_\mu)$ are the oscillation probabilities in vacuum. Provided
$a_{CP}(\delta = 0)$  is not too large.
\bea
a^{vac}_{CP} \approx a^{tot}_{CP} - a_{CP}(\delta = 0)
\label{asy}
\eea
is a good approximation. In any case, the error one makes in calculating
$a_{CP}$ by means of eq.(\ref{asy}) is smaller than the uncertainties on
$a_{CP}(\delta = 0)$ itself. In addition, the error can be estimated
by calculating the T-odd asymmetry \cite{kuo}, for each neutrino energy,
\bea\label{aTE}
a_{T}(E_\nu ,\delta ) = \frac{ P(\nu_e \rightarrow \nu_\mu) \;
\; -\; P(\nu_\mu \rightarrow \nu_e) }
{ P(\nu_e \rightarrow \nu_\mu) \;
\; +\; P(\nu_\mu \rightarrow \nu_e) }
\eea
where a nonzero value cannot be induced by matter effects. This also
means that $a_T$ a cleaner quantity in testing T violation than is
$a_{CP}$ for CP violation.

An important component of any study of muon appearance due to 
$\nu_e \rightarrow \nu_\mu$ oscillations is the event selection 
strategy for the $\mu$'s produced from charged current interactions
of the $\nu_\mu$'s. For neutrino experiments using a muon storage
ring, the detailed prescription for event selection can be formulated
only after the detector design is specified. There are, however,
some basic issues concerning the signal and the backgrounds which
all experiments are likely to be concerned with. 
On general grounds the background to a wrong sign muon signal is
associated with the numerous decay processes that can produce fake
muons: pions maskerading as muons, muonic charged currents (here one
would also have to miss the right sign muon) or electronic
charged currents (here one has also in the decay of the latter
right sign muons), to name only a few. Without referring to a specific
detector and the corresponding simulation toil we trust the
experimental proficiency by setting an overall detection 
efficiency of 30\% and by making a cut $E_\nu >$~5~GeV to eliminate
inefficiently observed low energy interactions.

\section{Who mixes, two, three, or four flavours?}

With growing evidence for non vanishing neutrino masses, experimental
studies of neutrino oscillations, and their analysis in terms of
three (or more) flavours, have become popular and will continue to be
of central significance for lepton physics in the future. 

The easiest way to describe any individual case of oscillations is to
use a scheme where only two neutrino flavours are allowed to
mix. Indeed, much work was done on analyses of neutrino oscillations
in terms of two flavours but, as was pointed out by us and by others, the
results for the squared mass differences may be misleading when
applied to the real lepton world which contains three flavours. We
summarize the situation regarding the squared mass differences as
follows. 

In models involving three oscillating flavours one often relies on
squared mass differences which are taken from analyses of individual
oscillation experiments in the framework of a two-flavour scheme. For
instance, if one assumes the solar, the atmospheric and the LSND
oscillations to be governed by just a single oscillation
``frequency'', $\Delta {\cal M}^2 $, then the characteristic
frequencies of the three oscillations, i.e.
\bea
\Delta {\cal M}^2 _{\mbox{solar}} 
= 10^{-10} \mbox{eV}^2\quad \mbox{or}\quad  10^{-5} \mbox{eV}^2
\nn \\
\nn \\
 10^{-3} \mbox{eV}^2 \leq \Delta {\cal M}^2_{\mbox{atmospheric}} \leq 
10^{-2} \mbox{eV}^2 \nn \\
\nn \\
10^{-1} \mbox{eV}^2 \leq \Delta {\cal M}^2_{\mbox{LSND}} \leq 
10^{1} \mbox{eV}^2 \nn
\eea
cannot be reconciled with just three neutrino mass eigenstates.
Therefore, in order to simultaneously accommodate all three
oscillations as observed,  under the assumption stated above, we must
introduce (at least) a fourth neutrino. Since we know from the width
of the $Z^0$~ boson that only three neutrino species have normal weak
interactions, this extra,  fourth neutrino must be sterile.

As there is no other, direct evidence for the existence of one or more 
sterile neutrinos, one is lead to conclude that assuming all observed
oscillation phenomena to involve but a single $\Delta {\cal M}^2  $ is
erroneous. Suppose, instead, that there are only three neutrinos, with
masses such that 
\bea\label{mass2diff}
m_3^2 -m_2^2 \equiv \Delta M^2 \ll m_2^2 -m_1^2 \equiv \Delta m^2
\eea
Then, as was shown in \cite{nos}, it is possible to explain the LSND result
as an oscillation involving $\Delta M^2$, the flavour conversion
of solar neutrinos as one involving $\Delta m^2 $ and the atmospheric
neutrino anomaly as a mixture of both frequencies. In contrast to
these findings, an analysis of the atmospheric data assuming
(erroneously) that only one $\Delta {\cal M}^2$ is involved would find
a value intermediate between those corresponding to the LSND and solar
effects, as observed.

In calculating observable effects of CP violation
in neutrino oscillations we assume a scenario with three flavours,
where the two squared mass differences obey the inequality
(\ref{mass2diff}) and lie in the range
\begin{equation}
  \label{values}
  10^{-4}\mbox{ eV}^2\le\Delta m^2\le 10^{-3}\mbox{ eV}^2\, ,\quad 
  \Delta M^2 \approx 0.3\mbox{ eV}^2\, .
\end{equation}

If there are three Dirac neutrino types, then the flavour eigenstates are
related to the mass eigenstates by a $ 3 \times 3$ unitary matrix
\bea\label{lCKM}
U=\pmatrix{c_{12}c_{13} & s_{12} c_{13} & s_{13} e^{-i \delta} \cr
-s_{12}c_{23}  - c_{12}s_{23}s_{13}e^{i \delta} & 
c_{12}c_{23}- s_{12}s_{23}s_{13}e^{i \delta} & s_{23}c_{13} 
\cr
s_{12}s_{23}- c_{12}c_{23}s_{13}e^{i \delta}  & 
- c_{12}s_{23}e^{i \delta} - s_{12}c_{23}s_{13} 
&c_{23}c_{13}}
\eea
where $c_{ij}=\cos\theta_{ij}$ and  $s_{ij}=\sin\theta_{ij}$. 
If the neutrinos are Majorana particles, there are two extra phases,
but these do not affect oscillations \cite{maj}.

The transition probability in vacuum for a neutrino changing from 
$\nu_i$ ($\bar{\nu}_i$) to $\nu_j$ ($\bar{\nu}_j$) is given by the sum 
and the difference of CP-even and CP-odd pieces, respectively,
\cite{lindner}
\bea
P(\nu_i \rightarrow \nu_j)& =& 
\PCPC(\nu_{i}\rightarrow\nu_{j}) + \PCPV(\nu_{i}\rightarrow\nu_{j}) \\
P(\bar{\nu}_{i}\rightarrow\bar{\nu}_{j}) &=&
\PCPC(\nu_{i}\rightarrow\nu_{j}) - \PCPV(\nu_{i}\rightarrow\nu_{j}),
\eea
where 
\bea
\PCPC(\nu_{i}\rightarrow\nu_{j}) &=& \delta_{ij} -4\mbox{Re}J^{ji}_{12}
\sin^2\Delta_{12} -4\mbox{Re}J^{ji}_{23} \sin^2\Delta_{23} -
4\mbox{Re}J^{ji}_{31} 
\sin^2\Delta_{31}, \nn \\
\\
\PCPV(\nu_{i}\rightarrow\nu_{j})
& = &-8\sigma_{ij} J \sin\Delta_{12} 
\sin\Delta_{23} \sin\Delta_{31}, \nn
\eea 
with $J$ the Jarlskog invariant and
\bea
J^{ij}_{kh} &\equiv &U_{ik} U_{kj}^\dagger
U_{jh} U_{hi}^\dagger \nonumber \\
\Delta_{ij}&\equiv &\Delta m^2_{ij}
L/4E \\
\sigma_{ij}&\equiv &\sum_k \varepsilon_{ijk} \nonumber
\eea

\section{Results for CP asymmetries with three flavours}

As stated above we henceforth assume a scenario with three flavours of
neutrinos characterized by the squared mass differences (\ref{values})
and the strong mixing found in \cite{nos}, solution I, where
\bea\label{angles}
\theta_{12}\approx 35.5^0\, ,\quad \theta_{23}\approx 27.3^0\, ,\quad
\theta_{13} \approx 13.1^0\; .
\eea

Although a detailed comparison might need further, refined analysis,
this range of squared mass differences and the set of mixing angles
(\ref{angles}) describes all observed neutrino anomalies in an overall 
and satisfactory manner. Here we show that this same set of parameters
predicts a CP asymmetry which may well be large enough to be
detectable with neutrino beams from a muon storage ring as described
in sect.~1.  

We organize the discussion of our results as
follows: We first present our main result for the expected
asymmetry. Next, the role of matter effects is illustrated by some
examples, followed by a comparison of CP violating asymmetries
with time reversal violating asymmetries. We then turn to a comparison 
with previous results and show why we find asymmetries which are
sizeably larger than the ones estimated previously. Finally, some
remarks on efficiencies and statistical uncertainties are added.

Let $n$ be the number of muons, $\overline{n}$ the number of antimuons 
detected in one year's time in a detector placed at 732~km from  the
collider. Assuming for a moment an efficiency of $100\%$, we find the 
asymmetry $(n-\overline{n})/(n+\overline{n})$ shown in Fig.~1, as a
function of the neutrino energy $E_\nu$. Part (a) of the figure refers 
to the lower limit $\Delta m^2 =10^{-4}$~eV$^2$,
part (b) refers to the upper limit $\Delta m^2 =10^{-3}$~eV$^2$ of
(\ref{values}). The solid line corresponds to setting $\delta =0$ (no
CP violation), the dashed line shows the full asymmetry, assuming
$\delta =\pi /2$. As a matter of example Fig.~2 shows the absolute
numbers $n$ and $\overline{n}$, obtained in one year of running, under 
the same assumptions as before and for $\Delta m^2 =10^{-3}$~eV$^2$. 

\noindent
\emph{The role of matter effects:} Clearly, the smaller matter effects 
the cleaner the measurement of the effects of genuine CP violation
will be. As the asymmetry due to charged-current matter interaction
grows faster than the CP asymmetry, as a function of the baseline,
intermediate distances between collider and neutrino detector are
preferred over long distances. We illustrate this observation
quantitatively by defining the asymmetry
\begin{equation}\label{asyCP}
  a_{CP}(E_\nu ,\delta ) = \frac{P(\nu_e\to\nu_\mu )- P(\bar{\nu}_e
    \to \bar{\nu}_\mu )} 
    {P(\nu_e\to\nu_\mu )+ P(\bar{\nu}_e
    \to \bar{\nu}_\mu )} 
\end{equation}
and by calculating the ratio $a_{CP}(E_\nu ,0)/[a_{CP}(E_\nu ,\pi /2)
- a_{CP}(E_\nu ,0)]$, as a function of $E_\nu$. Fig.~3 shows this
quantity for a baseline of 732~km, part (a), and a baseline of
7332~km, part (b), 
corresponding to the distance from FNAL to the Gran Sasso. In 
the case of the very long baseline, the CP asymmetry is completely
swamped by matter effects.

\noindent
\emph{CP- versus T-asymmetry:} We also computed the T-odd asymmetry
(\ref{aTE}) for the example $\Delta m^2 =10^{-4}$~eV$^2$ and the
shorter baseline $L=732$~km and compared it to the CP asymmetry
(\ref{asyCP}) from which the matter effects were subtracted. We found
them to agree within reasonable limits, thus corroborating the
approximation (\ref{asy}). 

\noindent
\emph{Comparison with previous results:} The authors of
ref.~\cite{belen} who use the following sample set of parameters
\begin{equation}\label{deRGH}
  \Delta m_{12}^2=10^{-4}\mbox{ eV}^2 ,\, \Delta
  m_{23}^2=10^{-3}\mbox{ eV}^2 , \, \sin^2\theta_{12} =.5 ,\,
  \theta_{23}=45^0 ,\, \theta_{13}=13^0\, ,
\end{equation}
find CP violating effects which are markedly smaller than the ones we
showed above. The explanation for this difference is simple: The most
noticeable difference between the set (\ref{deRGH}) and ours is the
value of $\Delta m_{23}^2$. Assuming the values (\ref{deRGH}) both the
(12)- and the 
(13)-channels are strongly affected by matter effects, the effective
sines $(\sin\theta_{12})_{matter}$ and $(\sin\theta_{13})_{matter}$ as 
defined in \cite{belen} quickly tend to 1 as the parameter $\mu$,
eq.~(\ref{matter}), increases. Consequently, the simultaneous
interplay of all three flavours and, hence, the visibility of CP
violation decrease. In contrast to this situation our values of
squared mass differences are such that only
$(\sin\theta_{12})_{matter}$ is affected while
$(\sin\theta_{13})_{matter}$ and, of course,
$(\sin\theta_{23})_{matter}$ remain unaffected. It is convenient to
define an effective mixing matrix $V_{ik}$ which is obtained from
(\ref{lCKM}) by replacing the sines and cosines by the matter affected 
sines and cosines, $(\sin\theta_{ik})_{matter}$, etc. 

To illustrate the
comparison Fig.~4(a) shows the pertinent, effective matrix elements
for our set of parameters, eqs.~(\ref{values}) and (\ref{angles}),
as a function of $\mu$, while Fig.~4(b) shows the same matrix
elements for the set (\ref{deRGH}). In the latter case, both $V_{11}$
and $V_{12}$ tend to zero with $\mu$ increasing to its value in the
Earth's crust.
 
\noindent
\emph{Efficiencies and statistical uncertainties:}
The discussion of statistical uncertainties is straightforward. 
In order to exclude the possibility that a measurement of a non vanishing
genuine CP violation is due to a statistical fluctuation, the
measured value must be larger than $n \cdot \delta (a_{CP})_{\mbox{\tiny
stat}}$, where $\delta (a_{CP})_{\mbox{\tiny
stat}}$ is the $1\sigma$ statistical error on $a_{CP}$ in the absence 
of CP violation, and $n$ is the number of standard deviations 
we require in order to be happy with our result. 
Since in absence of CP violation the expectations of
$n_{\nu_\mu}/N_{\mu^+}$ and $n_{\bar{\nu}}/N_{\mu^-}$ are equal, we get
\bea
\delta (a_{CP})_{\mbox{\tiny stat}} = \left( \frac{1}{4\langle
n_{\nu_\mu} \rangle} + \frac{1}{4\langle n_{\bar{\nu}_\mu} \rangle} 
\right)^{1/2}
\eea
where $ \langle n_{\nu_\mu} \rangle$ ($\langle n_{\bar{\nu}_\mu} \rangle $) is the
expected number of $\nu_\mu$ ($\bar{\nu}_\mu$) interactions seen in
the detector. 

Regarding the contribution of the background to the statistical
error, and according to the current estimates, the main source
of background will be due to charm production in the charged
current neutrino interactions in the detector \cite{juanjo}
\bea
\mu^- \rightarrow \nu_\mu \rightarrow &\mbox{CC interaction}&
\rightarrow \mu^- \mid_{\mbox{ 
\tiny lost}} \nn \\
c \rightarrow & \mbox{ c decay} &\rightarrow \mu^+  \mid_{\mbox{
\tiny found}} \nn
\eea 
Clearly, this background only affects the signal corresponding to the
$\bar{\nu_e} \rightarrow \bar{\nu}_\mu$ oscillations (because we
expect there  $\mu^+$ appearance), and therefore, this background 
``noise'' should be subtracted appropriately in the
counting of $n_{\bar{\nu}_\mu}$ in Eq.(\ref{stat}). Such a 
subtraction introduces a further source of statistical
error. Using the estimate \cite{juanjo} that 
\bea
n_{\bar{\nu}_\mu}\mid_{\mbox{\tiny back}} \simeq 10^{-5}
n_{\bar{\nu}_\mu}\mid_{P=1} 
\eea
where $n_{\bar{\nu}_\mu}\mid_{P=1}$ is the number of
antimuon neutrino interactions that would be
seen if all the initial antielectron neutrinos oscillated into
antimuon neutrinos, we find
\bea
\delta (a_{CP})_{\mbox{\tiny stat}} = \left( \frac{1}{4\langle
n_{\nu_\mu} \rangle} + \frac{1}{4\langle n_{\bar{\nu}_\mu} \rangle} 
+ \frac{10^{-5} \langle n_{\bar{\nu}_\mu}\mid_{P=1} \rangle
}{4\langle n_{\bar{\nu}_\mu} \rangle^2}
\right)^{1/2}
\eea
as the complete expression for the statistical error. It is important
to notice that even with only one year of data taking and a 
modest 30\% detecting efficiency, the statistical error will 
be small enough to rule out the possibility of attributing
to a statistical fluctuation a measurement
of a non vanishing CP violation.

\section{Conclusions}

In summary, the two rather different ``frequencies''
(\ref{mass2diff}), together with the strong mixing of all three
flavours that describe the solar neutrino deficit, the atmospheric
oscillations, and the LSND anomaly, lead to relatively large CP and
T~violating asymmetries in neutrino oscillations. With the set of
parameters (\ref{values}) and (\ref{angles}) the full interference of all
three flavours is well developed and is only moderately damped by
matter effects. We have also tried the other solutions to the neutrino 
anomalies we had found in \cite{nos} but find no more than 20\%
changes in the asymmetries. Among these the CP asymmetry seems large
enough to be measurable with neutrino beams from a 20~GeV muon
storage ring and with a detector at some 730~km from the source,
corresponding to the distance of the Gran Sasso laboratory from CERN
or, likewise, of the Soudan mine from FNAL.

\begin{center}
{\bf Acknowledgements}
\end{center}

We are very grateful to Karl Jakobs
for  enlightening
discussions. Financial support from the DFG is also acknowleged.

\newpage

\begin{figure}[!ht]
  \begin{center}
  \epsfig{file=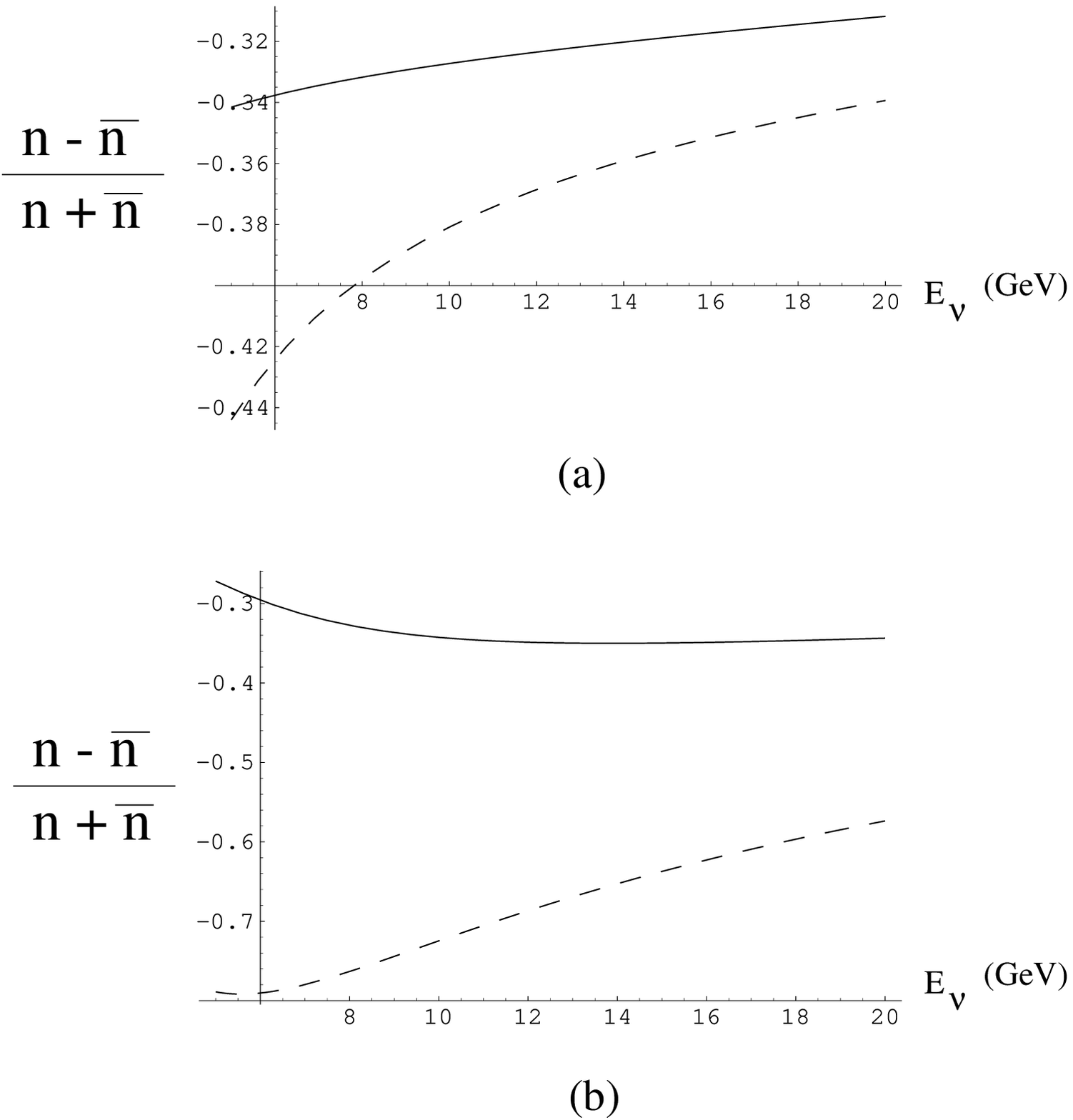,width=15cm}
\caption{$(n-\overline{n})/(n+\overline{n})$ as a function of the neutrino
energy for $\Delta m^2= 10^{-3}$ eV$^2$ (a) and 
$\Delta m^2= 10^{-4}$ eV$^2$ (b).  The solid line correspond to $\delta=0$
while the dashed line correspond to $\delta=\pi/2$ } 
  \end{center}
\end{figure}

\pagebreak

\begin{figure}[!ht]
  \begin{center}
  \epsfig{file=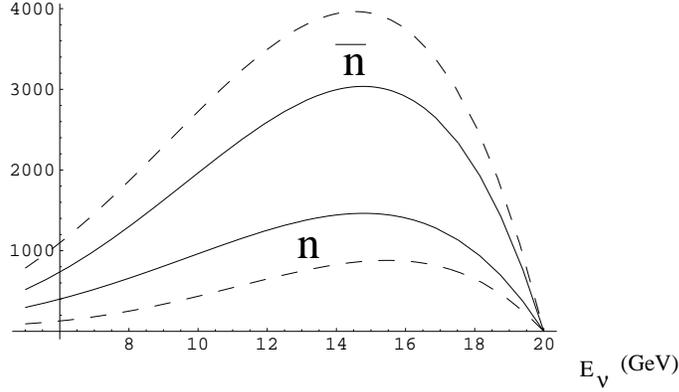,width=9cm}
  \parbox{15cm}{\caption{Absolute number of muons and antimuons detected in
one year's time in a 732 km baseline for  $\Delta m^2= 10^{-3}$ eV$^2$
and assuming 100\% detecting efficiency. dashed and solid lines as
before.
}}
  \end{center}
\end{figure}

\begin{figure}[!ht]
  \begin{center}
  \epsfig{file=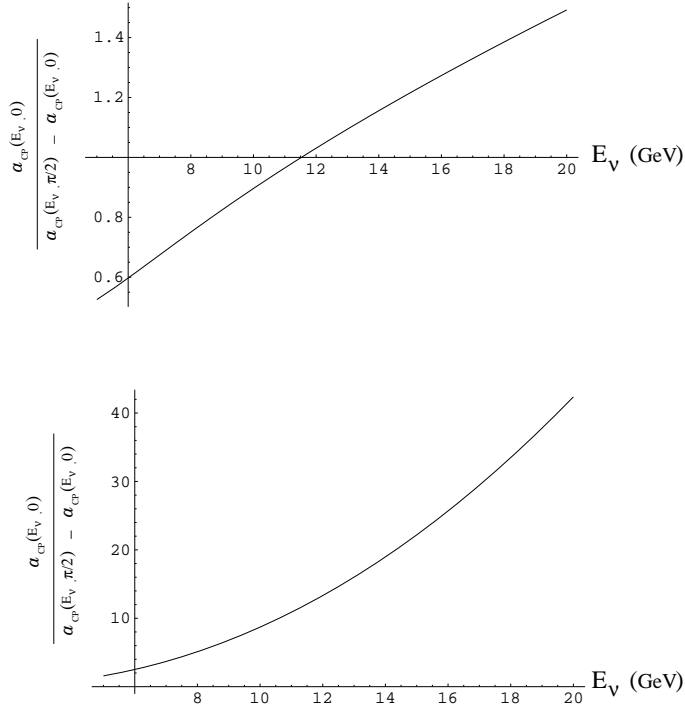,width=9cm}
  \parbox{15cm}{\caption{"Noise over signal" ratio as a function 
of the neutrino energy for $\Delta m^2= 10^{-3}$ eV$^2$ and a 
detector placed at 732 km (a) and 7332 km (b) from the collider.
}}
  \end{center}
\end{figure}

\begin{figure}[!ht]
  \begin{center}
  \epsfig{file=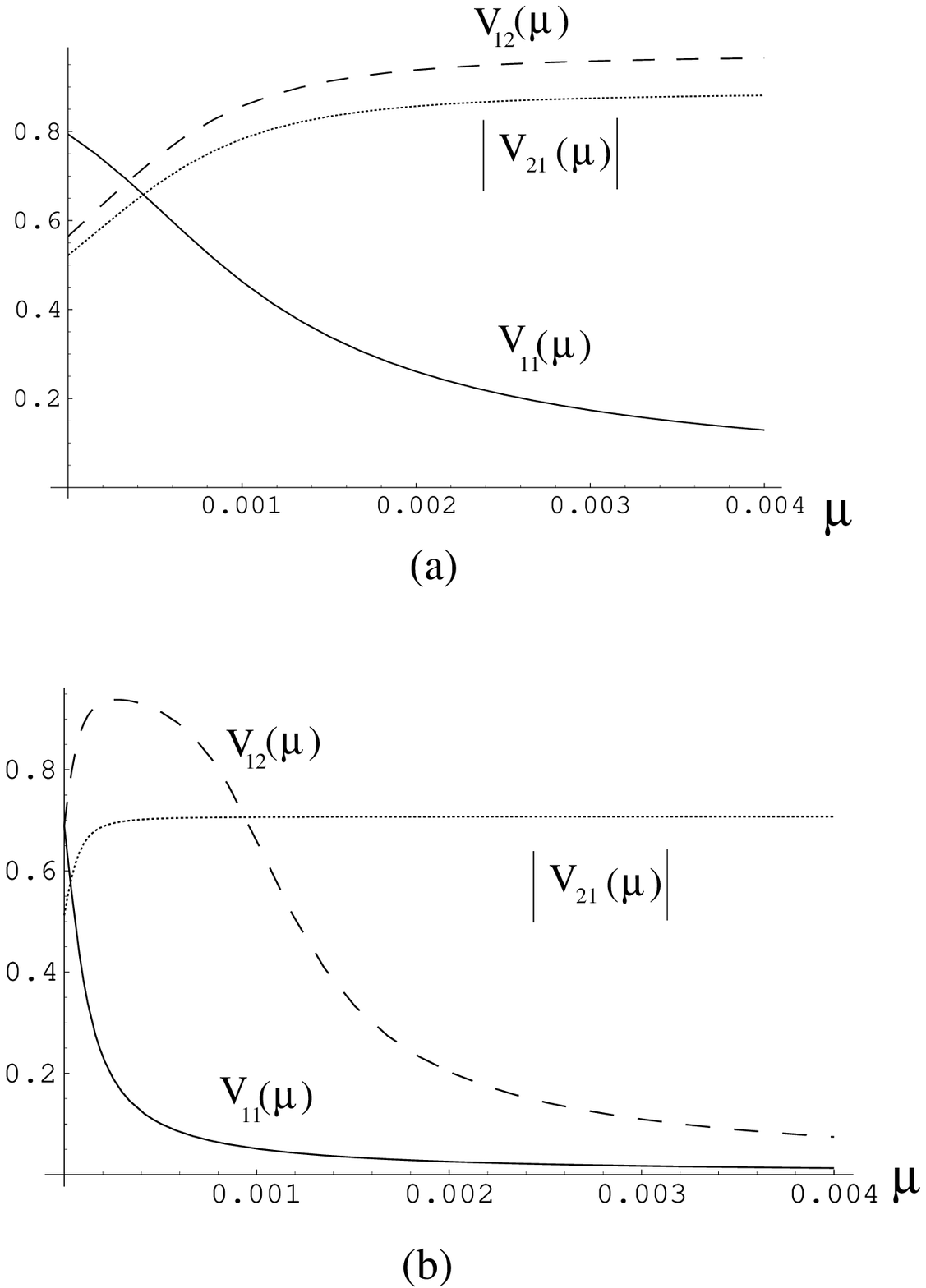,width=9cm}
  \parbox{15cm}{\caption{Effective matrix elements
as a function of the effective mass for the set of parameters used through
this work (a) and those of ref.~~\cite{belen}
}}
  \end{center}
\end{figure}


\vspace{.5cm}

\begin{thebibliography}{99}
\frenchspacing

\bibitem{FLC}
S. Geer, C. Johnstone, and D. Neuffer, FERMILAB-TM-2073;\\
D. Finley, S. Geer, and J. Sims, FERMILAB-TM-2072;\\
\emph{Prospective Study of Muon Storage Rings at CERN}, \\B. Autin,
A. Blondel, and J. Ellis (eds.), CERN~99-02/ECFA~99-197.

\bibitem{geer} 
S.H.Geer, \pr{D 57} (1998), 6989.

\bibitem{belen} A.deRujula, M.B.Gavela and P.Hernandez,
\np{B 547} (1999), 21.

\bibitem{belen2}
A.Donini, M.B.Gavela ,P.Hernandez and S.Rigolin, hep-ph/9909254.


\bibitem{rom}
A.Romanino, hep-ph/9909425.

\bibitem{lindner} 
K.Dick, M.Freund, M.Lindner and A.Romanino, hep-ph/9903308.


\bibitem{nosupdown} G. Barenboim and F. Scheck; \pl{B450} (1999), 189.

\bibitem{mat}
L.Wolfenstein, \pr{D 17} (1978), 2369;\\
S.P.Mikheyev and A.Y.Smirnov, Sov. J. \np{42} (1986) 913;\\
V.Barger {\it{et al.}}, \pr{D 22} (1980) 2718.

\bibitem{cab}
N.Cabibbo, \pl{B 72} (1978) 33.

\bibitem{ar}
J.Arafune, M.Koike and J.Sato, \pr{D 56} (1997), 3093\\
M.Tanimoto, \pl{B345} (1988) 373;\\
H.Minakata and H. Nunokawa, \pl{B 413} (1997) 369.

\bibitem{kuo}
T.Kuo and J.Pantaleone, \pl{B 198} (1987) 406.

\bibitem{maj} L.Wolfenstein, \pl{B 107} (1981), 77; \\
P.B.Pal and L. Wolfenstein, \pr{D 25} (1982), 766; \\
F.del Aguila and M.Zralek, \np{B 447} (1995), 211.


\bibitem{nos}
G.Barenboim and F.Scheck, \pl{B 440} (1998), 332.

\bibitem{juanjo}
A.Cervera, F.Dydak and J.J.Gomez-Cadenas, Nufact '99 Workshop,
Lyon, France.

\end{thebibliography}
\end{document}